\documentclass[useAMS,usenatbib,fleqn]{mn2e}
\usepackage{amsmath,amssymb, aas_macros,times}
\usepackage{graphicx}
\usepackage[usenames,dvipsnames]{xcolor}
\usepackage{color}
\usepackage{float}

\newcommand{\p}{^\prime}

\newcommand{\g}{\gamma}

\newcommand{\psim}{\lower.5ex\hbox{$\; \buildrel \propto \over\sim \;$}}
\newcommand{\lbar}{\lower.0ex\hbox{$\; \buildrel
{\lower0.0ex \hbox{-}} \over\lambda  \;$}}

\newcommand{\Msol}{{\rm M}M_\odot}

\newcommand{\km}{\mathrm{km}}
\newcommand{\erg}{\mathrm{erg}}

\newcommand{\s}{\mathrm{s}}

\newcommand{\Mpc}{\mathrm{Mpc}}
\newcommand{\Gpc}{\mathrm{Gpc}}
\newcommand{\yr}{\mathrm{yr}}
\newcommand{\Myr}{\mathrm{Myr}}
\newcommand{\Gyr}{\mathrm{Gyr}}

\newcommand{\ellE}{\ell_{\rm Edd}}

\newcommand{\medellE}{\tilde{\ell}_{\rm Edd}}
\newcommand{\pkellE}{\ell_{\rm Edd,pk}}

\newcommand{\pka}{a_{\rm pk}}

\begin{document}
\title{The Binary Black Hole Merger Rate from Ultraluminous X-ray Source Progenitors}





\author[Finke \& Razzaque]{
Justin D.\ Finke$^{1}$\thanks{justin.finke@nrl.navy.mil} and
Soebur Razzaque$^{2}$ \\ 
$^{1}$ U.S.\ Naval Research Laboratory, Code 7653, 4555 Overlook Ave.\ SW, Washington, DC, 20375-5352, USA \\
$^{2}$ Department of Physics, University of Johannesburg, PO Box 524, Auckland Park 2006, South Africa 
}

\maketitle

\begin{abstract}

Ultraluminous X-ray sources (ULXs) exceed the Eddington luminosity for
a $\approx 10M_\odot$ black hole.  The recent detection of black
  hole mergers by the gravitational wave detector ALIGO indicates
that black holes with masses $\ga 10M_\odot$ do indeed exist.
Motivated by this, we explore a scenario where ULXs consist of black
holes formed by the collapse of high-mass, low-metallicity stars, and
that these ULXs become binary black holes (BBHs) that eventually
merge.  We use empirical relations between the number of ULXs and the
star formation rate and host galaxy metallicity to estimate the ULX
formation rate and the BBH merger rate at all redshifts.  This assumes
the ULX rate is directly proportional to the star formation rate for a
given metallicity, and that the black hole accretion rate is
distributed as a log-normal distribution.  We include an enhancement
in the ULX formation rate at earlier epochs due to lower mean
metallicities.  With simplified assumptions, our model is able to
reproduce both the rate and mass distribution of BBH mergers in the
nearby universe inferred from the detection of GW 150914, LVT 151012,
GW 151226, and GW 170104 by ALIGO if the peak accretion rate of
ULXs is a factor $\approx$1 --- 300 greater than the Eddington
rate. Our predictions of the BBH merger rate, mass distribution, and
redshift evolution can be tested by ALIGO in the near future, which in
turn can be used to explore connections between the ULX formation and
BBH merger rates over cosmic time.

\end{abstract}

\begin{keywords}
gravitational waves --- X-rays: binaries --- stars: black
holes ---stars: massive --- accretion, accretion disks
\end{keywords}
\date{\today}
\maketitle

\section{Introduction}
\label{intro}

Ultraluminous X-ray sources (ULXs) are off-nuclear X-ray point sources
in nearby galaxies with X-ray luminosities greater than the Eddington
luminosity for a $\approx 10 M_\odot$ black hole
\citep[e.g.,][]{fabbiano88,mizuno99,colbert99,colbert02,liu05}.  They
are likely related to accretion onto high-mass black holes, although
their exact nature is still in question; until recently, there was no
reliable evidence for black holes with masses $\ga 10\ M_\odot$
outside of the supermassive black holes as the centers of galaxies.
It has been suggested that ULXs are a result of beamed, anisotropic
emission \citep*{georgan02,kording02}, although strong beaming
is now disfavored due to observations of ionized nebulae around some
ULXs that indicates their emission is isotropic
\citep[e.g.,][]{pakull03,gutierrez06,berghea10}.  The two remaining
possibilities are super-Eddington accretion
\citep[e.g.,][]{begelman01,poutanen07,finke07,sadowski15}, or
accretion onto black holes with masses $\ga 10 M_\odot$, perhaps even
as high as $10^4 M_\odot$, so-called intermediate-mass black holes
\citep[IMBHs;][]{colbert99}.  Hydrodynamic simulations of stellar
evolution and supernovae have indicated that such black holes could be
the remnants of very high mass, very low metallicity stars
\citep*[e.g.,][]{spera15}.  Indeed, there is some low significance
evidence that ULXs are preferentially found in metal-poor galaxies
\citep{mapelli10,prestwich13}.  Low-metallicity, Population III stars
were quite common in the early universe.  However, accretion from the
ISM onto IMBH remnants from the first stars are unlikely to be able to
explain the local ULX population \citep{volonteri05}.  A scenario
where the IMBH is produced by the collapse of a Population III star in
the early universe and then later captures a star through dynamical
interactions and becomes a ULX was shown by \citet{kuranov07} to be
incapable of explaining the ULX population.  Alternatively, high-mass
black holes could form through interactions in star clusters
\citep{downing11,morscher15} although in this case it is unlikely that
enough ULX could be formed by black holes capturing high-mass stars to
explain the local population \citep[e.g.,][]{blecha06}.

The detection of gravitational waves by the Advanced Laser
Interferometer Gravitational Wave Observatory (ALIGO) has given the
first evidence that black holes with masses $\ga 10 M_\odot$ exist.
Previous mass estimates $\ga 10 M_\odot$ of the black hole in IC 10
X-1 \citep{prestwich07,silverman08} are no longer considered reliable
\citep{laycock15,ligo16_astro}.  The first gravitational wave
  event detected, dubbed GW 150914, consisted of the merging of two
black holes with masses $36^{+5}_{-4}M_\odot$ and $29\pm4M_\odot$
merging to form a black hole with mass $62\pm4M_\odot$ at redshift
$z=0.09^{+0.03}_{-0.04}$ \citep{ligo16_astro,ligo16_detection},
although the pre-merger black hole mass estimates were later updated
to $35^{+5}_{-3}M_\odot$ and $30^{+3}_{-4}M_\odot$
\citep{abbott16_update}.  More merging black holes have been
  subsequently detected by ALIGO
  \citep{abbott16_gw151226,abbott16_observe1,abbott17}.  In this
paper, we re-evaluate the nature of ULXs in light of these
  discoveries.  We investigate a scenario for ULX formation similar
to that proposed by \citet{zampieri09}, that ULXs include high-mass
black holes formed by low-metallicity stars local to the ULX (i.e.,
not from stars in the early universe).  In our scenario, the ULXs are
born as binary high-mass field stars, before one of the stars explodes
in a supernova and turn into a black hole.  Thus, the black hole in a
ULX system is born not long before the ULX system was created.  The
extremely high mass stars needed to make the high-mass black holes do
indeed exist in the local universe; the $\eta$\ Carina system consists
of binary stars whose masses combine to be $250M_\odot$
\citep{kashi10}.  The fact that most high mass stars are found in
binaries \citep{garcia01,sana08} would also seem to make this scenario
likely.

In Section \ref{ULXform}, we develop a simple model for ULXs and
compute the ULX formation rate as a function of redshift.
\citet*{inoue16} have suggested that ULXs are the progenitors of BBH
mergers; that is, an accreting high-mass black hole and high-mass star
become a binary black hole (BBH) system after the companion star
explodes in a supernova; and the BBH eventually merges, and is
potentially detectable by ALIGO \citep[see also][]{bulik11}.  They use
this to estimate the rate of BBH mergers expected in the local
universe.  In Section \ref{BBHmerge} we use our knowledge of the ULX
formation rate to estimate the BBH merger rate.  Our calculation goes
beyond the estimates by \citet{inoue16} and explores detailed redshift
and metallicity dependences of the ULX and corresponding BBH merger
rates for different binary BH mass combinations.  Our modeling results
are compared with the rate and mass distribution inferred from mergers
that have been detected so far \citep[GW 150914, LVT 151012, GW
    151226, and GW 170104;][]{abbott16_observe1,abbott17}.  Finally,
we conclude with a discussion (Section \ref{discussion}).  In Appendix
\ref{metalenhance} we compute an enhancement in cosmological ULX
production due to decreasing metallicity at higher redshifts.

\section{ULX Formation}
\label{ULXform}

In this section, we discuss our simple ULX model, and
  its formation rate, both in the local universe and extending to high
  redshift.  We consider only those ULXs that are compact objects with
  high-mass companions, neglecting those with low-mass companions,
  which cannot become BBH systems.  We make a number of simplifying
  assumptions about the evolution of high mass stars, binaries, and
  supernovae.  We avoid discussion of poorly-understood tidal chemical
  mixing \citep[e.g.,][]{mandel16,demink16} and common envelope binary
  evolution \citep[e.g.,][]{belcz07,belcz16}.

\subsection{Simple ULX Model}

Consider two massive ($\ga 10M_\odot$), low-metallicity stars.  After
some time, one of the stars explodes in a supernova leaving a neutron
star or black hole remnant of mass $M_{\rm BH,1}$.  At some point in
the binary's life time, the the black hole is accreting from its
companion star, creating intense X-ray radiation with an X-ray
luminosity $L_X\ge10^{39}\ \erg\ \s^{-1}$.  The binary is in its ULX
phase.  The X-ray luminosity of the ULX can be expressed as a fraction
$\ell_{\rm Edd}$ of the Eddington luminosity, so that
\begin{flalign}
\label{LEdd}
L_X = \ell_{\rm Edd}\ L_{\rm Edd} = \ell_{\rm Edd}\ L_0\ m_{\rm BH,1}\ ,
\end{flalign} 
where $L_0=1.26\times10^{38}\ \erg\ \s^{-1}$ and $m_{\rm BH,1}=M_{\rm
BH,1}/M_\odot$. We assume that the ULX emission is isotropic,
consistent with observations of ionized nebulae around some ULXs.  We
take the ULX lifetime to be $t_{\rm ULX}=0.1$\ Myr, consistent with
several recent estimates \citep[see the discussion
by][]{mineo12,inoue16}.  However, this value is uncertain, as discussed
in Section \ref{discussion}.

Eventually, the massive companion star will exhaust its nuclear fuel
and also end its life in a supernova that leaves behind a compact
object of mass $M_{\rm BH,2}$.  We assume that $M_{\rm BH,2}\le M_{\rm
  BH,1}$ since generally lower mass stars live longer and make lower
mass compact objects than higher mass stars when in isolation,
although this can be altered in the mass exchange in binary
systems. We allow $m_{\rm BH,1}=M_{\rm BH,1}/M_\odot$ and $m_{\rm
  BH,2}=M_{\rm BH,2}/M_\odot$ to go down to 1.4, we actually are
including neutron stars as well, since some ULXs are known to be
accreting neutron stars
  \citep[e.g.,][]{bachetti14,israel17_ngc5907,israel17}.  However, we
will refer to them as black holes for most of this paper, since in our
model black hole ULXs will outnumber neutron star ULXs, consistent
with population synthesis modeling \citep{fragos15}, although if
  ULXs are anisotropic emitters, this may not be the case
  \citep{wikt17,middleton17}.  For the maximum mass of the compact
object, we use $m_{\rm BH,max}=130$ based on the maximum possible
black hole mass found from stellar evolution and supernova models of
low-metallicity stars by \citet{spera15}.

\subsection{Local ULX Formation}
\label{localULX}

We are interested in ULXs that eventually become BBHs, so these must
be ULXs with high-mass stellar companions.  There is evidence
that most ULXs in spiral galaxies have high-mass companions, while
ULXs in elliptical galaxies have low-mass companions
\citep[e.g.,][]{swartz04}.  We assume that all ULXs in spiral
galaxies have high-mass companions, and all ULXs in elliptical
galaxies have low-mass companions, so that we only explore those ULXs
that are found in spiral galaxies \citep[similar to][]{inoue16}.  In a
survey of nearby galaxies, \citet{walton11} found that the X-ray
luminosity function for ULXs in spiral galaxies is
\begin{flalign}
\label{Lfcn1}
N_X(L_X) = \frac{dN}{dL_X} \propto L_X^{-\alpha}
\end{flalign}
with $\alpha = 1.85\pm0.11$ and $L_X$ ranging from $L_{\min}=10^{39}\
\erg\ \s^{-1}$ to $L_{\max}=6\times10^{40}\ \erg\ \s^{-1}$.
Similarly, they find the number of ULXs per unit stellar mass of the
spiral host galaxy is 
\begin{flalign}
\label{Mgalfcn}
N_{X}(M_{Gal}) = \frac{dN}{dM_{Gal}} = N_0\ M_3^{-\beta}
\end{flalign}
where $M_{Gal} = 10^3\ M_3\ \Msol$ is the host galaxy
stellar mass\footnote{$\Msol=10^6M_\odot$}, $\beta = 0.64\pm0.07$, and
\begin{flalign}
N_0 = N_X(M_3=1) = 3.3\times10^{-4}\ \Msol^{-1}\ .
\end{flalign}
If the Milky Way has a stellar mass of
$(6.43\pm0.63)\times10^4\ \Msol$ \citep{mcmillan11} then Equation
(\ref{Mgalfcn}) predicts $\approx1.5$ ULXs in our Galaxy,
approximately consistent with 0 that are observed.  Combining
Equations (\ref{Lfcn1}) and (\ref{Mgalfcn}) we expect that
\begin{flalign}
\label{Lfcn2}
N_X(L_X, M_{Gal}) & = \frac{dN}{dL_X dM_{Gal}} 
= N\p_0\ M_3^{-\beta}\ L_{40}^{-\alpha}\ ,
\end{flalign}
where $L_{40} =
  L_X/(10^{40}\ \erg\ \s^{-1})$ and $N\p_0 =
  N_X(L_{40}=1,M_3=1)$. Here we assume that $N_X(L_X)$ and
  $N_X(M_{Gal})$ are independent. Using
\begin{flalign}
N_X(M_{Gal}) = \int_{L_{\min}}^{L_{\max}}\ dL\ N_X(L_X, M_{Gal})
\end{flalign}
one can solve for the normalization constant in Equation (\ref{Lfcn2}), 
\begin{flalign}
N\p_0 = 
 2.4\times10^{-5}\ \Msol^{-1}\ (10^{40}\ \erg\ \s^{-1})^{-1}\ .
\end{flalign}
The ULX formation rate as a function of $L_X$ and host
galaxy mass can be estimated as
\begin{flalign}
\label{ULXformrate}
\dot{N}_X(L_X, M_{Gal}) & = 
\frac{dN}{dtdL_XdM_{Gal}} 
\nonumber \\
& = \frac{N_X(M_{Gal}, L_X) }{t_{\rm ULX}}\ .
\end{flalign}

The cosmological ULX formation rate per unit comoving volume, per unit
X-ray luminosity, in the local universe can be found by convolving
$\dot{N}_X(L_X, M_{Gal})$ with the spiral galaxy
mass function in the local universe, $\phi(M_{Gal}; z=0) =
dN/(d(\log_{10}M_{Gal})dV)$, so that 
\begin{flalign}
\label{ULXratez0}
\dot{n}_X(L_X; z=0) & = \frac{dN}{dVdtdL_X}
\nonumber \\ &
= \int_{M_{Gal, \min}}^{M_{Gal, \max}} dM_{Gal}\ 
\nonumber \\ & \times
\frac{\phi(M_{Gal}; z=0)}{\ln(10)}\ 
\nonumber \\ & \times
\dot{N}_X(L_X, M_{Gal})\ 
\end{flalign}
where we use $M_{Gal,\min}=10^2\ \Msol$ and $M_{Gal,\max}=10^7\
\Msol$.  The function $\phi(M_{Gal}; z=0)$ was found by
\citet{moffett16}, which they represented by a Schechter function,
\begin{flalign}
\label{schechter}
\phi(M_{Gal}; z=0) & = \phi^*\ln(10) \left(\frac{M_{Gal}}{M^*}\right)^{1+\delta}
\nonumber \\ & \times
\exp\left( - \frac{M_{Gal}}{M^*} \right)\ .
\end{flalign}
They found it was well-fit with parameters
$\phi^*=(8.55\pm1.00)\times10^{-4}\ \Mpc^{-3}$, $\delta=-1.39\pm0.02$,
and $M^*=10^{10.70\pm0.05}M_\odot$, which we use.  The total ULX
formation rate density for all ULXs can be found by integrating,
\begin{flalign}
\label{ndotxtotz0}
\dot{n}_{X,tot}(z=0) = \int^{L_{\max}}_{L_{\min}} dL_X\ \dot{n}_X(L_X;z=0)\ .
\end{flalign}

\subsection{Cosmological ULX Formation}

There is evidence that high-mass X-ray binaries are connected with
star formation \citep[e.g.,][]{grimm03,ranalli03}.  Since in our model
ULXs are formed from short-lived, high mass stars, the ULX formation
rate should follow the star formation rate.  However, in our model
ULXs are formed from low-metallicity stars, and the average
metallicity decreases with redshift.  There is in fact some weak
evidence that lower metallicity galaxies host more ULXs for a given
star formation rate \citep[e.g.,][]{mapelli10,prestwich13}.  In
Appendix \ref{metalenhance} we use some of this evidence to estimate a
metallicity enhancement in ULX formation, $\zeta(z)$.

Using the knowledge of the SFR, metallicity enhancement, and local ULX
formation rate as found in Section \ref{localULX}, one can find the
ULX formation rate density for any $z$,
\begin{flalign}
\label{nXz}
\dot{n}_X(L_X;z) & = \frac{\psi(z)}{\psi(z=0)} \zeta(z)
\ \dot{n}_X(L_X;z=0)\ ,
\end{flalign}
where $\psi(z)=dM_*/(dVdt)$ is the star formation rate
(SFR) density.  \citet*{finke10_EBL} found that the combination of the
\citet{cole01} SFR, with parameters found by \citet{hopkins06},
combined with an initial mass function (IMF) from \citet{baldry03},
reproduced the luminosity density data available at the time better
than other combinations of SFR and IMF.  Therefore, we use this
$\psi(z)$ here. 

\begin{figure}
\vspace{10.0mm} 
\includegraphics[width=8cm,height=12cm,keepaspectratio]{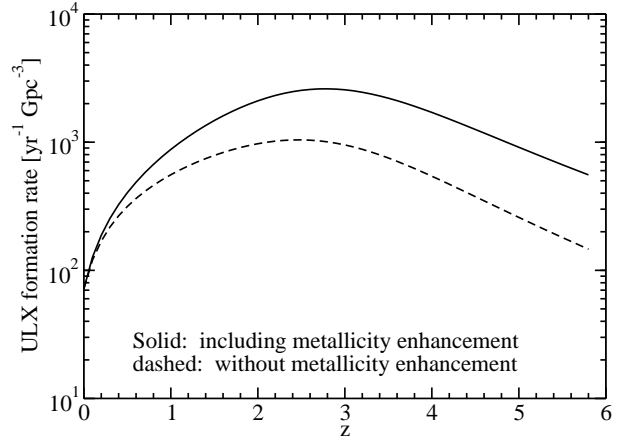}
\caption{Total ULX formation rate density $\dot{n}_{X,tot}(z)$
computed from Equation (\ref{ndotxtot}), with and without the
metallicity enhancement factor, $\zeta(z)$.  }
\label{ULXformationrate}
\vspace{2.2mm}
\end{figure}

The total ULX formation rate density for all ULXs can be found by
integrating,
\begin{flalign}
\label{ndotxtot}
\dot{n}_{X,tot}(z) = \int^{L_{\max}}_{L_{\min}} dL_X\ \dot{n}_X(L_X;z)\ ,
\end{flalign}
where recall $L_{\min}=10^{39}\ \erg\ \s^{-1}$ to
$L_{\max}=6\times10^{40}\ \erg\ \s^{-1}$. This result is
shown in Figure \ref{ULXformationrate}, with and without the
metallicity enhancement factor (i.e., the latter has $\zeta(z)=1$).
Not surprisingly, the ULX formation rate follows the shape of the star
formation rate without the metallicity enhancement; including the
enhancement, the ULX formation rate is larger at higher
redshifts.

\section{BBH Mergers}
\label{BBHmerge}

\subsection{Model}

After the companion star ends its life in a supernova, and joins its
companion as a black hole, the BBH system is created.  
Thus far, we have made no assumptions about the Eddington ratio of the
ULX.  However, to use the ULX formation rate to estimate the BBH
merger rate, we must use a distribution of Eddington ratios to convert
from the ULX X-ray luminosities to the BBH masses.
A study of accreting stellar-mass black holes indicates that this
distribution resembles a log-normal distribution, 
\begin{flalign}
\label{Pell}
P_\ell(\ellE) & = \frac{A_\ell}{\ellE}
\\ \nonumber & \times
\exp\left\{ \frac{-[\ln(\ellE)-\ln(\pkellE)]^2-\sigma^2}{2\sigma^2}\right\}\ ,
\end{flalign}
with peak $\pkellE\approx10^{-2.5}-10^{-2}$ and $\sigma\approx 1$
\citep{reynolds13}. A study of X-ray selected broad-line active
galactic nuclei by \citet{suh15} found the Eddington ratios of these
objects are distributed as a log-normal distribution, with
$\pkellE=10^{-0.6}$ and $\sigma=0.8$.  If the accretion
  flow of ULXs behaves similarly to accreting stellar-mass black holes
  and active galactic nuclei, it seems reasonable to assume that this
  distribution of $\ellE$ for ULXs is also a log-normal
  distribution. Here we will use a log-normal distribution
for $P_\ell(\ellE)$ with $\sigma=1$ and various values for $\pkellE$,
which we consider a free parameter.  Since we have
$L_{\max}=6\times10^{40}\ \erg\ \s^{-1}$, and the maximum possible
black hole mass was found to be $130 M_\odot$ by \citet{spera15}, the
Eddington ratio must extend to at least $\ell_{\rm Edd}>3.7$.  Thus,
we will allow the value of $\ellE$ to be $>1$ to explain the brightest
ULXs, although the log-normal distribution may be more
  questionable for ULXs with $\ellE\gg 1$, since the distribution was
  found for objects with lower $\ellE$.  In equation
(\ref{Pell}), the normalization constant is found by performing the
integral
\begin{flalign}
1 = \int_{\ell_{\min}}^{\ell_{\max}} d\ellE\ P_\ell(\ellE)
\end{flalign}
and solving for $A_\ell$, where $\ell_{\min}=L_{\min}/(L_0 m_{\rm
BH,1})$ and $\ell_{\max}=L_{\max}/(L_0 m_{\rm BH,1})$. 

The BBHs' orbits decay by emitting gravitational waves, and they merge after
\begin{flalign}
\label{tGW1}
t_{\rm GW} & = 5.6\times10^7\ \left( \frac{a}{10 R_\odot}\right)^4
\left( \frac{M_{\rm BH,1}}{30M_\odot}\right)^{-1}\ 
\nonumber \\ & \times
\left( \frac{M_{\rm BH,2}}{30M_\odot}\right)^{-1}\ 
\left( \frac{M_{\rm BH,1}}{30M_\odot}+\frac{M_{\rm BH,2}}{30M_\odot}\right)^{-1}\ 
\yr\ 
\end{flalign}
\citep{peters64} where $a$ is the initial separation of
the BBHs. If a binary black hole system is created at
redshift $z_{X}$ and merges at $z_m$, then
\begin{flalign}
\label{tGW2}
t_{\rm GW} = \int^{z_{X}}_{z_m} dz \left| \frac{dt}{dz} \right|
\end{flalign}
where 
\begin{flalign}
- \frac{dt}{dz} = \frac{1}{H_0(1+z) \sqrt{ (1+z)^3\Omega_m + \Omega_\Lambda} }\ 
\end{flalign}
in a flat $\Lambda$CDM universe.  To find $z_X$ for a given $z_m$,
$m_{\rm BH,1}$, $m_{\rm BH,2}$, and $a$, we
solve Equations (\ref{tGW1}) and (\ref{tGW2}) numerically using
standard cosmological parameters $H_0=70\ \km\ \s^{-1}\ \Mpc^{-1}$,
$\Omega_m=0.3$, and $\Omega_\Lambda=0.7$.  Doing this
requires knowledge of the distributions of $m_{\rm BH,2}$, and
$a$. 

\citet{abbott16_observe1} use a power-law distribution for $m_{\rm
BH,2}$ and we follow their calculation, so that the distribution,
normalized to unity, is
\begin{flalign}
P_m(m_{\rm BH,2}) = A_mm_{\rm BH,2}^\g
\end{flalign}
where
\begin{equation}
A_m = \left\{\begin{array}{ll}
(\g+1)/(m_{\rm BH,1}^{\g+1} - m_{\rm BH,min}^{\g+1}) & \g \ne 1 \\
1/\ln(m_{\rm BH,1}/m_{\rm BH,min}) & \g = 1 
\end{array}
\right. \ .
\end{equation}
 \citet{abbott16_observe1} use $\g=0$, and consequently so do we,
although $\g$ is in principle a free parameter. 

The separation of binaries can change during their lives due to
  mass exchange between the two stars due to Roche lobe overflow and
  winds, tidal interactions, magnetic braking, and emission of
  gravitational radiation.  Further, when massive stars explode the
  resulting compact object can get a significant ``kick'', which can
  substantially alter the binary separation.  These processes are
  modeled in detail in sophisticated population synthesis codes
  \citep[e.g.,][]{hurley02,belczy02,belczy08}.  For our purposes, we
  are only interested in the separation when binaries become BBHs.
  Population synthesis calculations by \citet{belczy02} indicate that
the initial separation of the BBHs is a distribution with a peak of
about $a=10R_\odot$ in their ``standard model''.
Therefore, we use a log-normal distribution for the initial separation
of the BBHs, normalized to unity over $a=0$ to
$a=\infty$,
\begin{flalign}
P_a(a) & = \frac{1}{\sigma_a a \sqrt{2\pi}}
\\ \nonumber & \times
\exp\left\{ \frac{-[\ln(a)-\ln(\pka)-\sigma_a^2]^2}{2\sigma_a^2}\right\}\ ,
\end{flalign}
with peak $\pka=10R_\odot$ and $\sigma_a=0.6$.  
In general, however, $\pka$ can be considered another free
parameter.

Putting everything together, the merger rate density 
\begin{flalign}
\label{ndotm}
\dot{n}_m(m_{\rm BH,1};z_m) 
& = \frac{dN}{dV dt dm_{\rm BH,1}}
\nonumber \\ 
& =  \Biggr\{ \int_{\ell_{\min}}^{\ell_{\max}} d\ellE\ P_\ell(\ellE) 
\nonumber \\ & \times
\frac{L_X}{m_{\rm BH,1}} \dot{n}_X(L_X; z=0) \Biggr\}
\nonumber \\ & \times
\int_{m_{\rm BH,min}}^{m_{\rm BH,1}} dm_{\rm BH,2}\ P_m(m_{\rm BH,2})\ 
\nonumber \\ & \times
\int_0^{\infty} da\ P_a(a)
\nonumber \\ & \times
\frac{\psi(z_X(z_m,m_{\rm BH,1},m_{\rm BH,2},a))}{\psi(z=0)} 
\nonumber \\ & \times
\zeta(z_X(z_m,m_{\rm BH,1},m_{\rm BH,2},a))\ ,
\end{flalign} 
where $\ell_{\min}=L_{\min}/(L_0 m_{\rm BH,1})$,
$\ell_{\max}=L_{\max}/(L_0 m_{\rm BH,1})$, $L_X/m_{\rm
BH}$ and $dL_X/dm_{\rm BH}$ are calculated from Equation (\ref{LEdd}),
and $\dot{n}_X(L_X;z=0)$ is given by Equation (\ref{ULXratez0}).  The
total merger rate
\begin{flalign}
\label{ndotmtot}
\dot{n}_{m,tot}(z_m) = \int_{m_{\rm BH,min}}^{m_{\rm BH,max}}\ dm\ \dot{n}_m(m;z_m)\ ,
\end{flalign}
where recall $m_{\rm BH,min}=1.4$ and $m_{\rm BH,max}=130$.

\begin{figure}
\vspace{10.0mm} 
\includegraphics[width=8cm,height=12cm,keepaspectratio]{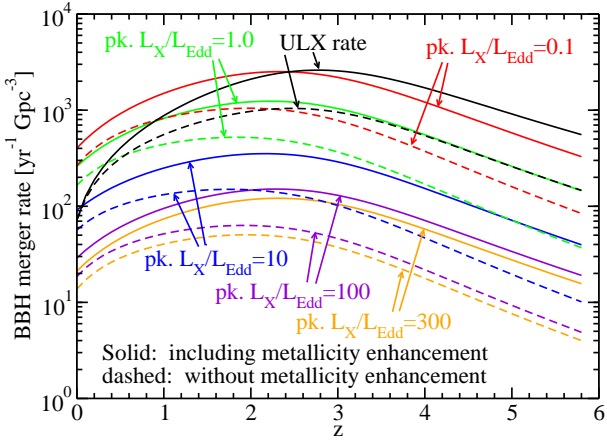}
\caption{Total BBH merger rate density, $\dot{n}_{m,tot}(z_m)$,
computed from Equation (\ref{ndotmtot}), with and without the
metallicity enhancement factor, $\zeta(z)$.  Different values of the
peak Eddington ratio, $\pkellE$, were used,
as indicated on the plot.  The ULX formation rate, as in Figure
\ref{ULXformationrate}, is also shown as indicated.} 
\label{mergerrate}
\vspace{2.2mm}
\end{figure}

The total merger rate is plotted in Figure \ref{mergerrate}, again,
with and without the metallicity enhancement factor $\zeta(z)$.  The
overall shape of the curves roughly follows the ULX formation rate,
delayed by a time $t_{\rm GW}$.  The lower the Eddington ratio
$\pkellE$, the more BBH mergers will originate from the observed
ULX progenitor population.

\begin{figure}
\vspace{10.0mm} 
\includegraphics[width=8cm,height=12cm,keepaspectratio]{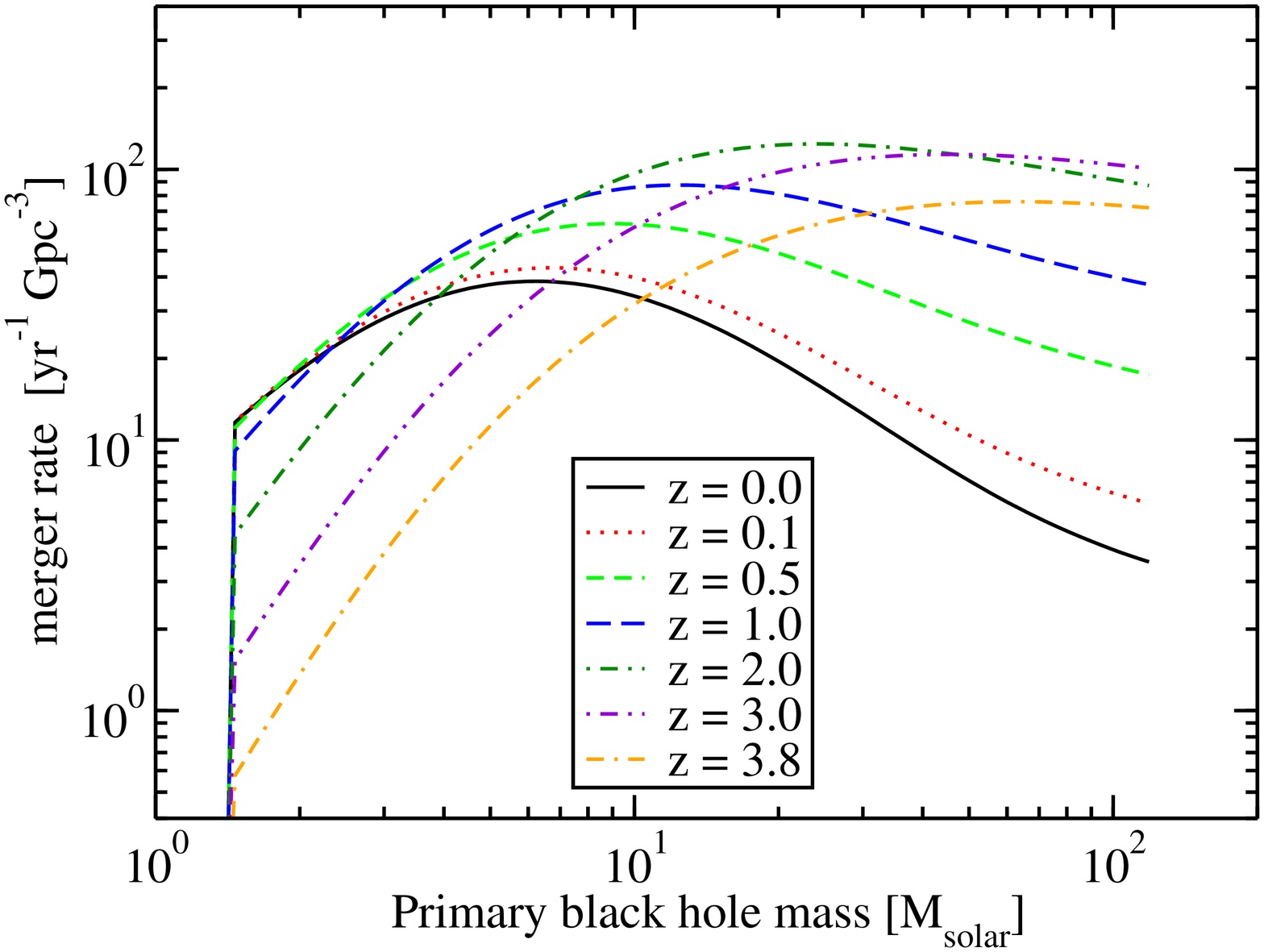}
\caption{BBH merger rate density per unit logarithmic primary
black hole mass bin, $m_{\rm BH,1}\times\dot{n}_m(m_{\rm BH,1};z_m)$,
plotted as a function of primary black hole mass, $m_{\rm BH,1}$, for
various values of $z_m$.  This calculation uses $\pkellE=10$\ and
includes the metallicity enhancement factor.  }
\label{massspectrum}
\vspace{2.mm}
\end{figure}

\begin{figure}
\vspace{3.0mm} 
\includegraphics[width=8cm,height=12cm,keepaspectratio]{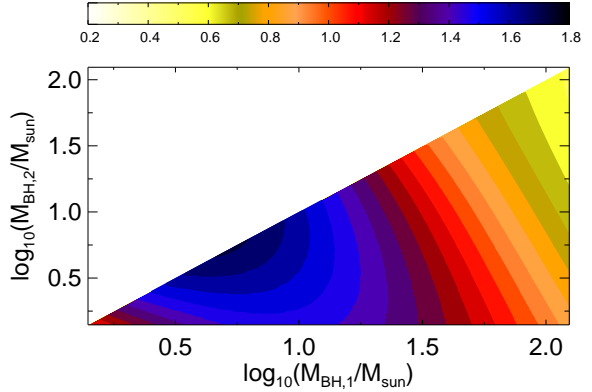}
\caption{BBH merger rate density per unit logarithmic black hole
mass bin plotted as a function of $m_{\rm BH,1}$ and $m_{\rm BH,2}$,
for $z_m=0$.  This calculation uses $\pkellE=10$\ and includes the
metallicity enhancement factor.  The contours show $\log_{10}[m_{\rm
BH,1}\dot{n}_m(m_{\rm BH,1};z_m=0)/(\yr^{-1}\ \Gyr^{-3})]$ as
indicated by the bar.}
\label{massspectrumcontour}
\vspace{2.mm}
\end{figure}

The BBH merger rate per unit logarithmic primary black hole mass
bin, $m_{\rm BH,1}\times \dot{n}_m(m_{\rm
BH,1};z_m)$, for $\pkellE=10$\ is plotted as a function of
$M_{\rm BH,1}$ for various values of $z_m$ in Figure
\ref{massspectrum}.  At low values of $M_{\rm BH,1}$ mergers are
fewer, since there is not enough time for many of them to occur within
the age of the universe.  At higher redshifts, low-mass mergers become
fewer in number, since the universe is younger and there is less time
for mergers to occur.  Thus, overall the peak increases at high $z$.
This means this model produces few binary neutron star systems such as
PSR 1913+16 \citep*{taylor79} and PSR J0737$-$3039 \citep{burgay03}
that will merge within the age of the universe.  However, we note that
the population synthesis calculations of \citet{belczy02} find a
separate peak at $a=0.3R_\odot$ for the initial separation of binary
neutron stars.  So another population, which will not go through a ULX
phase, can explain the binary neutron stars.  Also note the overall
normalization increases up to $z\approx 3$ and decreases at higher
$z$, in agreement with Figure \ref{mergerrate}.  In Figure
\ref{massspectrumcontour} $m_{\rm BH,1}\times
\dot{n}_m(m_{\rm BH,1};z_m=0)$\ is plotted as a function of
$m_{\rm BH,1}$ and $m_{\rm BH,2}$.  This was computed from Equation
(\ref{ndotm}) using $P_m(m_{\rm BH,2})\rightarrow\delta(m_{\rm
BH,2}-m\p_{\rm BH,2})$.  The peak in merger rate is where $m_{\rm
BH,1}=m_{\rm BH,2}\approx6$, in agreement with Figure
\ref{massspectrum}.

\begin{figure}
\vspace{10.0mm} 
\includegraphics[width=8cm,height=12cm,keepaspectratio]{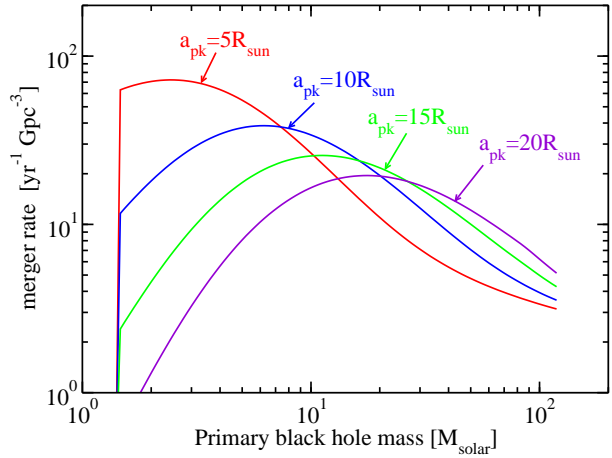}
\caption{BBH merger rate density per unit logarithmic primary black hole 
mass bin, $m_{\rm BH,1}\times\dot{n}_m(m_{\rm BH,1};z_m=0)$, plotted
as a function of primary black hole mass, $M_{\rm BH,1}$, for
$\pkellE=10$, plotted for various values of the initial BBH merger
separation, $a$, as indicated on the plot.  This calculation includes
the metallicity enhancement factor. }
\label{massspectrum_comparea}
\vspace{2.5mm}
\end{figure}

The BBH merger rate per unit logarithmic primary black hole mass
bin at $z_m=0$ as a function of $m_{\rm BH,1}$ is plotted in Figure
\ref{massspectrum_comparea} for different values of the peak initial
BBH separation $\pka$, instead of our ``default'' value
$\pka=10R_\odot$.  Clearly, both the low-mass cutoff, and peak are
strongly dependent on $\pka$ This leads us to the conclusion that
future observations of mergers by ALIGO could build up enough
statistics to observe this enhancement, which would then constrain the
initial BBH separation.

\begin{figure}
\vspace{10.0mm} 
\includegraphics[width=8cm,height=12cm,keepaspectratio]{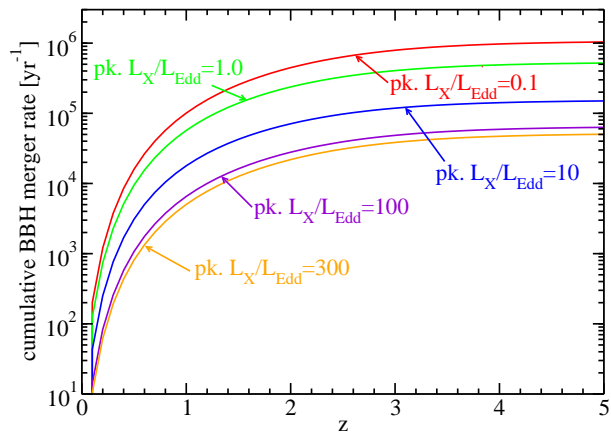}
\caption{Cumulative BBH merger rate density, $\dot{N}_{m,tot}(<z)$,
computed from Equation (\ref{cumrate}), including the metallicity
enhancement factor.}
\label{mergerratecumulative}
\vspace{2.2mm}
\end{figure}

\begin{figure}
\vspace{3.0mm} 
\includegraphics[width=8cm,height=12cm,keepaspectratio]{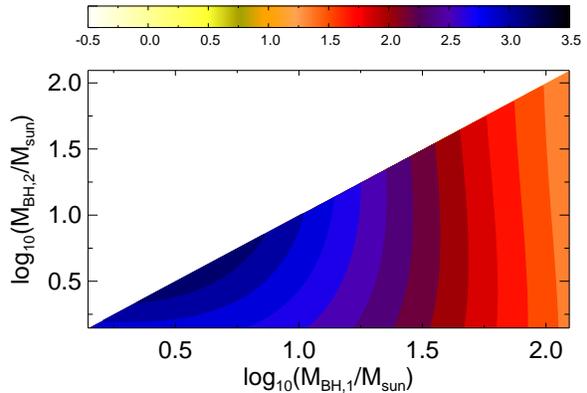}
\caption{Cumulative BBH merger rate for $z<1$, plotted as a function
of $M_{\rm BH,1}$ and $M_{\rm BH,2}$.  This calculation uses
$\pkellE=10$\ and includes the metallicity
enhancement factor.  The contours show
$\log_{10}[\dot{N}_{m,tot}(z<1)/(\yr^{-1})]$ as indicated by the bar.}
\label{cumulativecontour}
\vspace{2.mm}
\end{figure}

The cumulative merger rate as measured by the observer can be found from
\begin{flalign}
\label{cumrate}
\dot{N}_{m,tot}(<z) = 4\pi\int_0^{z}\ dz_m\ \frac{\dot{n}_{m,tot}(z_m)}{1+z}\ \frac{dV_c}{dz_m}
\end{flalign}
where
\begin{flalign}
\frac{dV_c}{dz} = \frac{c\ d_c(z)^2}{H_0 \sqrt{ (1+z)^3\Omega_m + \Omega_\Lambda} }\ ,
\end{flalign}
and
\begin{flalign}
d_c(z) = \frac{c}{H_0}\int_0^z \frac{dz\p}{\sqrt{ (1+z\p)^3\Omega_m + \Omega_\Lambda} }\ 
\end{flalign}
is the comoving distance.  This is plotted in Figure
\ref{mergerratecumulative} for models that include our metallicity
enhancement factor.  The cumulative BBH merger rate
follows the same basic pattern as the merger rate (Fig.\
\ref{mergerrate}), with lower $\pkellE$ models having higher merger
rates. The cumulative merger rate for up to $z=1$,
approximately the limiting redshift of ALIGO when it reaches its
design sensitivity, as a function of $m_{\rm BH,1}$ and
$m_{\rm BH,2}$ is shown in Figure \ref{cumulativecontour}.  This was
again calculated with $P_m(m_{\rm BH,2})\rightarrow\delta(m_{\rm
BH,2}-m\p_{\rm BH,2})$.  Mergers are expected to be maximized at
$m_{\rm BH,1}\approx m_{\rm BH,2}\approx 6$.

\subsection{Comparison with Other BBH Merger Models}

\citet{inoue16} also used ULXs to estimate the merger rate in the
local universe.  Our calculations generally predict a merger rate
greater than theirs (see their Figures 1 and 2).  This
  is despite the fact that our luminosity function is actually
  slightly lower than the one they use by a factor of 2, below the
  exponential cutoff they use at high $L_X$.  This is due to our
  taking into account the changing star formation rate, which is
  greater at higher $z$.

Our merger rate estimation can also be compared with the calculations
by \citet{mandel16}.  Those authors simulated the formation of BBH
systems from chemically homogeneous evolution of closely-interacting
massive binary stars.  All our models shown in Figure
\ref{mergerrate} are above theirs.  The evolution at higher redshift
is quite different.  \citet{mandel16} find a peak of $\approx$\ 20
Gpc$^{-3}$\ yr$^{-1}$ at $z\approx0.5$, then a merger rate declining
sharply after that, so that there are hardly any mergers above
$z=1.5$; see their Figure 7.

By contrast, estimates by \citet{dominik13} find large numbers of BBH
mergers ($\ga10-100\ \yr^{-1}\ \Gpc^{-3}$) out to redshift $z\approx
16$, mainly depending on whether or not they include ``high kicks''.
Their ``high kick'' scenario seems to be ruled out by rate inferred by
the detection of BBH mergers by ALIGO, since this scenario predicts
$\la 1\ \yr^{-1}\ \Gpc^{-3}$ at low $z$.  In their scenarios without
high kicks, they find 10--100 $\yr^{-1}\ \Gpc^{-3}$ at $z=0$, 
consistent with our models with $\pkellE\ga1.0$.
Their results for the evolution of the merger rate at high $z$ is also
different from ours; while our merger rates peak at $z=2-3$, in all of
their scenarios the merger rate peaks at $z\ga4$.

The predictions for the evolution of BBH merger rate with $z$ by
\citet{mandel16} and \citet{dominik13} not only differ substantially
from our predictions, but differ substantially from each other as
well.  Both these authors take a more ``first principles'' approach,
while our calculation is more empirically-motivated.  We hope this
makes it more accurate, although we note that there are many unknowns
in our calculation (discussed further in Section \ref{discussion}).
ALIGO will be able to detect BBH mergers out to no higher than
$z\approx1$ when it reaches its peak sensitivity in $\sim$2020
\citep{ligo16_astro}; the differences in the merger rate predictions
by ourselves, \citet{mandel16}, and \citet{dominik13} at higher $z$
are unlikely to be observed in the near future.  However, the striking
rapid evolution in the merger rate with $z$ predicted by
\citet{mandel16} at $z<1$ should be confirmed or refuted with further
ALIGO observations.

\subsection{Comparison with ALIGO Observations}
\label{ALIGOcompare}

Based on the ALIGO detection of three BBH mergers: GW 150914, GW
151226, the lower significance LVT 151012, \citet{abbott16_observe1}
constrained the BBH merger rate to be between 9 and 240
Gpc$^{-3}$\ yr$^{-1}$\ in the local universe, with the most likely
value being $\approx 70\ \Gpc^{-3}\ \yr^{-1}$ with an event-based
analysis, and $\approx 100\ \Gpc^{-3}\ \yr^{-1}$ with a power-law fit
to the distribution of $m_{\rm BH,1}$ (more on this below).  The
  additional detection of GW 170104 tightens this range a bit, to
  12--213 Gpc$^{-3}$\ yr$^{-1}$ \citep{abbott17}.  All of our model
curves in Figure \ref{mergerrate} consistent with $\pkellE>1.0$ are
within this range, with the closest model being the one with $\pkellE=
10$.  We conclude, based on the inferred ALIGO merger rate, that our
model is a good representation if $\pkellE\ga 1.0$ for ULXs.

\begin{figure}
\vspace{10.0mm} 
\includegraphics[width=8cm,height=12cm,keepaspectratio]{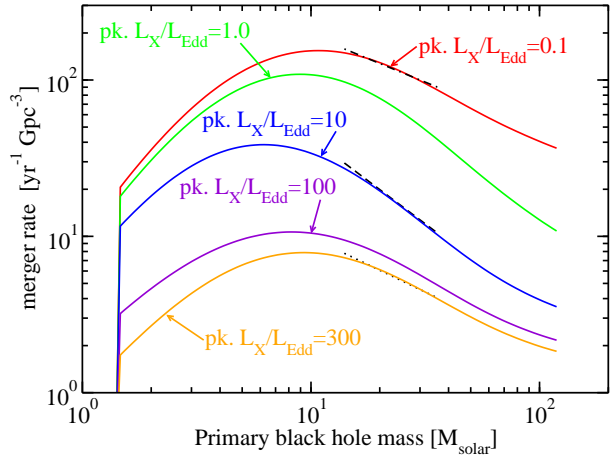}
\caption{BBH merger rate density, $m_{\rm BH,1}\times\dot{n}_m(m_{\rm
BH,1};z_m=0)$, plotted as a function of primary black hole mass,
$m_{\rm BH,1}$, for various values of the median Eddington ratio,
$\medellE$, as indicated on the plot.  The black dashed, dotted, and
dashed-dotted lines show power-laws, $\dot{n}_m(m_{\rm
BH,1};z_m)\propto m_{\rm BH,1}^{-2.1}$, $\propto m_{\rm BH,1}^{-1.7}$,
and $\propto m_{\rm BH,1}^{-1.6}$, respectively, between
$m_{\rm BH,1}=14$ and $m_{\rm BH,1}=36$, the range of primary black
hole masses observed by ALIGO. }
\label{massspectrum_compare}
\vspace{2.mm}
\end{figure}

From ALIGO detections of BBH mergers, one can infer more information
than just the total merger rate.  Based on detections in ALIGO's
  first observing run, and assuming that $\dot{n}_m(m_{\rm
  BH,1};z_m\approx0)\propto m_{\rm BH,1}^{-\theta}$,
\citet{abbott16_observe1} infer $\theta=2.5^{+1.5}_{-1.6}$,
for mergers with $m_{\rm BH,1}$ between $14.2^{+8.3}_{-3.7}$ (GW
151226) and $36.2^{+5.2}_{-3.8}$ (GW 150914).  With the
  addition of GW 170104, this was revised to
  $\theta=2.3^{+1.3}_{-1.4}$. In our model, on a plot
of the merger rate as a function of $m_{\rm BH,1}$, the slope is
weakly dependent on $\pkellE$, as Figure \ref{massspectrum_compare}
demonstrates.  Here we also plot various power-laws consistent with
the inferred ALIGO distribution for $14<m_{\rm BH,1}<36$.  The mass
distributions for all of our models shown in Figure
\ref{massspectrum_compare} are consistent with the ALIGO observations,
which have admittedly large error bars.  Our model is thus fairly
robust with respect to the free parameter $\pkellE$, and we conclude
that our model with $\pkellE\approx 1-300$ can reproduce both the
inferred rate and mass distribution of BBH mergers.

\section{Discussion}
\label{discussion}

We have explored the scenario where ULXs consist of a massive black
hole and companion star, become BBH systems, where the BBHs finally
merge and create gravitational waves potentially detectable by ALIGO.
By assuming that the ULX formation rate is directly proportional to
the star formation rate (modulo metallicity effects that are taken
into account as described in Appendix \ref{metalenhance}), we computed
the ULX formation rate as a function of redshift.  We then assumed
that every ULX becomes a BBH, and took into account the time for a BBH
merger to occur, to compute the BBH merger rate.  As discussed in
  Section \ref{ULXform}, we made a number of simplifying assumptions
  about the evolution of high mass stars, binaries, and supernovae.
  For instance, \citet{dominik13} have a fairly large suite of compact
  object merger rate models with various assumptions about whether
  Hertzsprung Gap stars can be common-envelope donor stars, whether or
  not delayed supernovae occur, and other variations.

Our derivation of the metallicity enhancement factor (Appendix
\ref{metalenhance}) makes use of a number of tentative results:
namely, the relation between the metallicity of a host galaxy and the
enhancement in the number of ULXs it produces, and in the evolution of
the average metallicity of galaxies with redshift.  For the latter, we
note that a similar correlation from \citet{kewley05,kewley07} was
used by \citet{mandel16}; and it is very similar to the ``high-end''
metallicity evolution used by \citet{dominik13}.  Further studies on
the relationship of the number of ULXs and host galaxy metallicity,
and the evolution of metallicity in the universe, could greatly
improve our estimates.

We find that our model is consistent with the rate and distribution of
BBH mergers inferred from the ALIGO detections if $\pkellE=1-300$. Our
scenario requires super-Eddington accretion on to black holes with
masses $\le130M_\odot$. This means that if our model is accurate and
the BBH mergers observed by ALIGO originated as ULXs, a large
fraction, possibly even the great majority, of ULXs have
super-Eddington accretion.  There is evidence for super-Eddington
accretion in some ULXs; for example the discovery of pulsar ULXs
requires super-Eddington accretion 
  \citep{bachetti14,fuerst16,israel17_ngc5907,israel17}.  The X-ray
spectrum of a ULX in NGC 5907 in various states led \citet{walton15}
to conclude that this source was accreting at a super-Eddington rate.
ALIGO could see 100s of BBH mergers in the near future
\citep[e.g.,][]{demink16}.  As it detects more mergers, the large
errors on the inferred rate and distribution will get smaller, and our
model parameters, such as $\pkellE$ and $\pka$ could be constrained.

One major unknown is the ULX timescale, $t_{\rm ULX}$.  We use the
value $t_{\rm ULX}=0.1$\ Myr for all ULXs, the value used by
\citet{mineo12} and \citet{inoue16}.  This value is consistent
with the thermal timescale of high-mass main sequence stars.  
  \citet{king01} considered the thermal timescale to be the most
  likely relevant timescale for the lifetime of a ULX.  However, the
companion stars of ULXs may not be main sequence stars; indeed 11/62
ULXs examined by \citet{heida14} show infrared excesses consistent
with being red supergiants; spectroscopic follow-up for one of these
indicates it is indeed a red supergiant \citep{heida15}.  Red
supergiants have considerably shorter thermal timescales, and thus
ULXs with red supergiant companions would have considerably shorter
lifetimes.  If all sources had shorter $t_{\rm ULX}$, our predicted
merger rate would be lower for a given $\pkellE$, and higher $\pkellE$
would be required to be consistent with the observed BBH merger rate
from ALIGO.  However, most high-mass X-ray binaries do not have
supergiant companions, and if this is also true for ULXs with high
mass companions (the only ones we are interested in here), on average
ULXs would have $t_{\rm ULX}\sim0.1$\ Myr \citep{mineo12}. 
  thus we use this value.  If the nuclear burning timescale
  is the relevant timescale for the lifetime of a ULX, the timescale
  could be as long as $t_{\rm ULX}=1\ \Myr$ \citep{patruno08}. There
  is some evidence for these longer $t_{\rm ULX}$ from the kinematics
  of the nebula around the ULX NGC 1313 X-2 \citep{pakull02}.  These
  longer lifetimes are consistent with the ages of the star clusters
  where ULXs reside, which have been constrained to be $\la$4--20 Myr
  \citep{ramsey06,grise11,poutanen13}.  If we had used a larger
  lifetime, $t_{\rm ULX}=1\ \Myr$, lower values of
$\pkellE$ would be consistent with the ALIGO constraints on the merger
rate.

If a large fraction of ULXs are neutron stars, their
  X-ray emission could be beamed.  This could alter our results; see
  e.g., \citet{king01}; \citet{wikt17}, \citet{king17}; or
  \citet{middleton17} for an exploration of this possibility.  If
  beaming decreased the actual luminosity a fraction $b$ relative to
  the observed luminosity, there would be a factor $1/b$ more sources
  that are not seen.  Thus, the acutal number of ULXs would be a
  factor $1/b$ higher.  If $b$ is inversely proportional to accretion
  rate or intrinsic X-ray luminosity \citep[see e.g.][]{king09} then
  naturally the enhancement would be greater for brighter sources.
  The effect of a $b<1$ on the total BBH merger rate is less obvious,
  but we expect it would alter the distribution of masses (e.g.,
  Fig.\ \ref{massspectrum}) so that there would be more low-mass
  mergers and fewer high-mass mergers.  \citet{israel17_ngc5907} found
  that a beaming factor $b\approx 1/7$ for the neutron star ULX in NGC
  5907 is consistent with expectations for a thin disk and avoidance
  of the ``propeller'' mechanism.  

It is possible that BBHs occur from dynamical interactions in star
clusters, and that this channel dominates the BBH merger rate 
  \citep[e.g.,][]{sadowski08,oleary16,chatterjee17}.  Since we
neglect BBH mergers that occur from other formation channels, and
since not every BBH may originate as a ULX, the BBH merger rates we
compute could be viewed as lower limits.  This is also true because we
neglect mergers that originated as accreting binaries with
luminosities $L_X<10^{39}\ \erg\ \s^{-1}$, as described at the end of
Section \ref{BBHmerge}.  However, again, our model reproduces the
inferred rate and distribution of BBH mergers from ALIGO 
  \citep{abbott16_observe1,abbott17}, so that mergers from other
origins are not needed.  It has been suggested that the
  spin orbit alignment of the detected mergers make a dynamical origin
  for them more likely \citep{rodriguez16,abbott17}.  However, a
  population synthesis simulation that incorporates stellar rotation
  indicates that an isolated binary evolution scenario can explain the
  ALIGO measurements \citep{belcz17}.  The distribution of masses
  observed by ALIGO (i.e., $\theta$ in Section \ref{ALIGOcompare})
  could be an important clue to determining whether the binary
  evolution of dynamic channel dominates the formation of BBH mergers.
  In the coming years, we will learn much about BBHs and
ULXs through the detection of gravitational waves.

\section*{Acknowledgements}

J.D.F.\ would like to thank Teddy Cheung, James Steiner, Michael
Wolff, and Kent Wood for useful discussions.  We are grateful to
the anonymous referee for helpful comments that have improved this
paper.  J.D.F.\ was supported by the Chief of Naval Research.
S.R. was partially supported by the National Research Foundation
(South Africa).

\appendix

\section{Metallicity Enhancement Factor}
\label{metalenhance}

\begin{figure}
\vspace{12.0mm} 
\includegraphics[width=7cm,height=12cm,keepaspectratio]{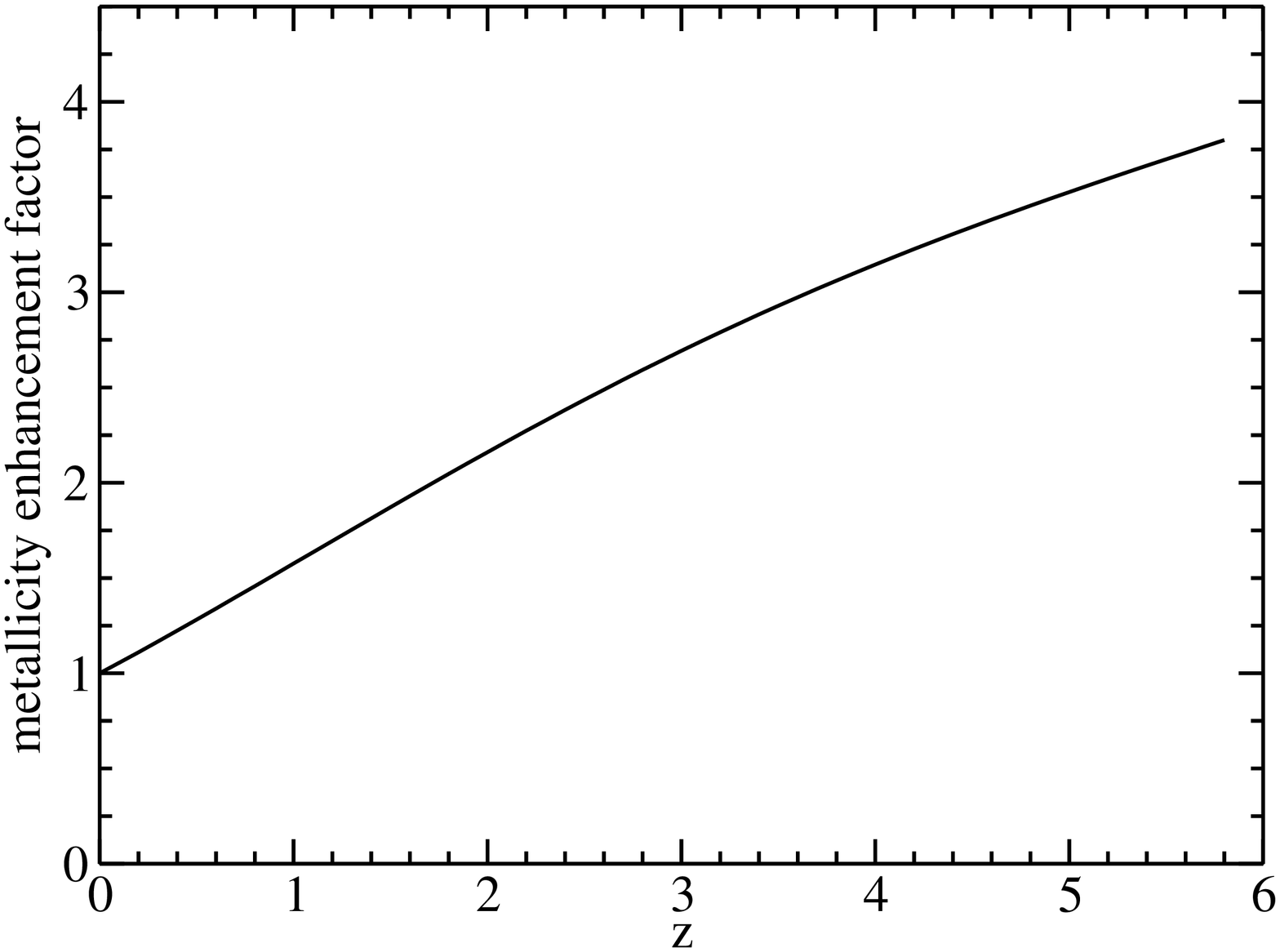}
\caption{The metallicity enhancement factor, $\zeta(z)$.}
\label{zetaplot}
\vspace{2.mm}
\end{figure}

A relationship between galaxy stellar mass and metallicity was found
by \citet{tremonti04} using a study of 53,000 galaxies at $z\sim0.1$
from the Sloan Digital Sky Survey.  \citet{ma16} use cosmological
simulations to derive the galaxy mass-metallicity relation and its
evolution with redshift, with the simulation results matching
observations at $z=0-3$.  We use the relationship from \citet{ma16},
which is
\begin{flalign}
\label{massmetal}
F_{\rm OH}  & =  4.45 + 0.35\log_{10}(M_{Gal}/M_\odot) + 0.93\exp(-0.43z)
\end{flalign}
where we define $F_{\rm OH} \equiv 12+\log_{10}({\rm O/H})$, and ${\rm
O/H}$ is the average gas-phase oxygen to hydrogen number ratio in the
galaxy.  A study by \citet{mapelli10} found that the number of ULXs
per star formation rate in a sample of spiral, irregular, and peculiar
galaxies was anti-correlated with the galaxies' metallicities.  Taking
the solar value to be $F_{\rm OH,\odot}=8.81$, their result can be
written as an enhancement factor $A_{\rm OH}$ in ULX formation
relative to the solar value as
\begin{flalign}
\log_{10}[A(F_{\rm OH})] = 4.85 - 0.55 F_{\rm OH}\ .
\end{flalign}
The significance of this correlation was low, but at present it seems
to be the best that can be done.  We then define a metallicity
enhancement as a function of redshift by
\begin{flalign}
\label{zeta}
\zeta(z) & \equiv \frac{ \int \frac{dM_{Gal}}{M_{Gal}} \phi(M_{Gal};z) A(F_{\rm OH}) }
{ \int \frac{dM_{Gal}}{M_{Gal}} \phi(M_{Gal};z)  }\ 
\nonumber \\ & \times
\frac{\int \frac{dM_{Gal}}{M_{Gal}} \phi(M_{Gal};z=0)}
{\int \frac{dM_{Gal}}{M_{Gal}} \phi(M_{Gal};z=0) A(F_{\rm OH}) }\ ,
\end{flalign}
where $F_{\rm OH}$ is found from $M_{Gal}$ from Equation
(\ref{massmetal}).  This function $\zeta(z)$ gives the enhancement in
ULX formation per SFR at redshift $z$, integrated over all galaxies at
that redshift, relative to the ULX production rate at $z=0$.  Here the
galaxy stellar mass function as a function of $z$ from the GOODS
survey is used \citep{fontana06}.  It is still parametrized as a
Schechter function, Equation (\ref{schechter}), but this time the
parameters as a function of $z$ are given by
\begin{flalign}
\phi^* & = 0.0035(1+z)^{-2.2}\ \Mpc^{-3},
\\
\log_{10}(M^*/M_\odot) & = 11.16 + 0.17z - 0.07z^2\ ,
\\
\delta & = -1.18 -0.82z\ .
\end{flalign}
In Equation (\ref{zeta}) all integrals are performed from $10^8\
M_\odot$ to $10^{13}\ M_\odot$, the same values used by
\citet{fontana06} to calculate the total stellar mass as a function of
$z$.  The calculation of $\zeta(z)$ is plotted in Figure
\ref{zetaplot}.

\vspace{1mm}
\bibliographystyle{mn2e}
\bibliography{references,ULX_ref,gravwave_ref,EBL_ref,mypapers_ref}

\end{document}